\documentclass[manuscript]{acmart}

\usepackage[utf8]{inputenc}
\usepackage{lscape}
\usepackage{rotating} 
\usepackage{algorithm} 
\usepackage{algpseudocode} 
\usepackage{subcaption}  
\usepackage{xcolor}

\usepackage{multirow}
\usepackage{graphicx}

\usepackage[normalem]{ulem}
\useunder{\uline}{\ul}{}

\usepackage{adjustbox}
\usepackage{hyperref}

\AtBeginDocument{%
  \providecommand\BibTeX{{%
    \normalfont B\kern-0.5em{\scshape i\kern-0.25em b}\kern-0.8em\TeX}}}

\setcopyright{acmcopyright}
\copyrightyear{2026}
\acmYear{2026}
\acmDOI{XXXXXXX.XXXXXXX}

\acmConference[Conference acronym 'XX]{Make sure to enter the correct
  conference title from your rights confirmation email}{June 03--05,
  2018}{Woodstock, NY}
\acmISBN{978-1-4503-XXXX-X/18/06}

\begin{document}

\title[Fairness Explanations Improve Fairness In Human-AI Decision Making]{Fairness Explanations Improve Fairness In Human-AI Decision Making: an empirical evaluation of parole decision-making}


\author{Peter Daish}
\email{peter.daish@swansea.ac.uk}
\affiliation{%
  \institution{Computational Foundry, Swansea University}
  \city{Swansea}
  \country{UK}
}

\author{Alan Dix}
\affiliation{
  \institution{Cardiff Metropolitan University}
  \city{Cardiff}
  \country{UK}
}

\author{Muneeb Ahmad}
\affiliation{
\institution{Computational Foundry, Swansea University}
\city{Swansea}
\country{UK}
}

\author{Matt Roach}
\affiliation{
\institution{Computational Foundry, Swansea University}
\city{Swansea}
\country{UK}
}

\renewcommand{\shortauthors}{Daish et al.,}

\ccsdesc[500]{Human-centered computing~Empirical studies in HCI}

\keywords{fairness, explainable AI, human-AI decision-making, bias mitigation}

\begin{abstract}
Both humans and AI exhibit systemic biases, yet little is known about how interface design can foster fairness in hybrid human-AI decision-making. This paper introduces a method for extracting fair data samples from biased datasets and adapts a fairness-explanation technique to highlight potential AI bias during decisions. We conducted a within-subjects study (n=48) where non-experts made decisions across three conditions: independent human judgment, unexplained AI assistance, and AI assistance augmented with visual ethnic-bias explanations (Fair-XAI). To isolate effects, the AI conditions provided systematically balanced recommendations. Results show that fairness explanations significantly reduced both sex and ethnic bias compared to human-only decisions. Fair-XAI also marginally improved fairness over balanced AI recommendations while enhancing task performance. These findings demonstrate that fairness-centered explanation interfaces can meaningfully shape user behavior, providing concrete design insights for developing equitable and accountable human-AI systems.
\end{abstract}
\maketitle

\section{Introduction}

Prior work has shown that people exhibit biases when making judgements~\cite{doubling_back_bias,janis_space_shuttle_group_think,titan_group_think,Biernat_Vescio_2002,notre_dame_law_review_unconscious_bias_ethnic,Miller_2019}, AI systems reproduce biases~\cite{accuracy_equality,dwork_fairness_2012,hardt_equality_2016,angwin_machine_2016} and that AI may train humans to learn biases~\cite{humans_inherit_bias}. Whilst prior research efforts have explored the mechanisms through which biases are represented in human and AI decisions, there is a marked gap in knowledge regarding \textit{how human awareness to AI biases influences decision-making outcomes}. This paper investigates whether raising human awareness to AI bias can improve the fairness of decision-making through human-AI teams.

Much of prior literature explores biases in human and AI decision-makers \textit{independently}: research agenda within AI fairness domains typically centres on exposing and mitigating biases picked up during the AI training process~\cite{mehrabi_survey_2022, Caton_Haas_2024}, whilst retaining utility~\cite{accuracy_fairness_tradeoff}. By contrast, fairness as a property of human decision-makers has been typically studied descriptively, with little by way of the production of empirically evaluated guidelines or techniques that promote fairer decision-making practices~\cite{Fasolo_Heard_Scopelliti_2025}. When it comes to the combination of humans and machines, research agenda has been dominated by an emphasis on improving utility, not fairness~\cite{bansal_does_2021,vivian_lai_towards}.

Following suit, the advent of `explainable AI' technology -- interpretable tools that approximate blackbox AI behaviours -- has resulted in the development of new methods that enable people to explore utility-, as opposed to fairness-centric, influences over AI decision outcomes~\cite{ribeiro2016lime,lundberg2017shap,xai_feature_importance_popularity}. Even among explanation techniques designed with the express desire to convey AI-fairness qualities to human users, empirical evaluations typically only explore human perceptions of the fairness of AI decision-makers~\cite{Srivastava_Heidari_Krause_2019, dodge_iui_rebuttal} rather than assessing the decision-making performance of the human-AI teams themselves. Consequently, there is a \textit{distinct} gap in knowledge regarding how human-AI decision-makers interact to exacerbate or potentially, mitigate, biases. Neither does there exist guidelines nor empirical insights on how human-AI interactive environments may be designed to promote principles of fairness being enacted through human-AI teams. This is important, as expert social decision-domains -- such as healthcare and the judiciary -- where AI systems are known to pose a risk to health~\cite{race_bias_clinical_decisions,allocation_of_covid_funding_race_bias,hidden_in_plain_sight_race_correcting_algorithms} and social justice~\cite{angwin_machine_2016} are rapidly adopting AI technologies and pairing them with human decision-makers without empirical guidance regarding whether human-AI teams help safeguard social decisions or else drive inequalities.

In this paper we address these gaps by introducing a viable visual fairness explanation designed to communicate model bias through an explainable AI lens. We evaluate our explanation's ability to change fairness, accuracy, confidence, automation bias, and usability in a parole decision task by means of a within-subject study (n=48) with non-expert decision-makers. Participants performed decision-making across three, order counterbalanced, experimental conditions: unassisted human judgement, AI-assisted, fairness explained AI-assisted. To our knowledge, this is the \textit{first study} to evaluate fairness explanations as an XAI technique that explicitly communicates social bias relative to an external normative standard for the purpose of mitigating bias at the point of decision-making.

This paper makes three contributions. First, it presents a design prototype and algorithmic methodology for generating visual fairness explanations to foreground AI bias to human decision-makers. Second, it presents a methodology for generating a bias-controlled dataset from a historically biased source, helping isolate bias arising from human decision-makers from bias inherited from the data. Third, it reports the results of an empirical study evaluating whether fairness explanations improve fairness in human-AI decision-making.


\begin{itemize}
    \item \textbf{RQ1:} How does explicit knowledge of potential AI ethnic bias, communicated through visual fairness explanations at the point of decision-making, affect human-AI parole judgments?
    \begin{itemize}
     \item \textbf{RQ1a:} Does it improve fairness and accuracy?
    \item \textbf{RQ1b:} Does it reduce automation bias?
    \item \textbf{RQ1c:} What effect does it have on usability?
    \end{itemize}
\end{itemize}

\section{Related Work}

\subsection{Decision--Making: Individuals, Groups and AI}

Human decision--making can suffer under increased workloads and is often susceptible to influence from external factors. These factors can include: emotional response and affective state (mood influenced by external factors rather than the task at hand)~\cite{emotional_social_pressure_effect_decision_making, mood_emotion_decision_making}, social pressure~\cite{emotional_social_pressure_effect_decision_making}, in addition to cognitive~\cite{cognitive_biases_decision_making, cognitive_bias_four_types_crisis_decisions} and social biases caused by stereotypes~\cite{green_implicit_bias_clinical_decisions_2007, mood_stereotyping_bias}. Prior work has also found that group and collective dynamics are effective in reducing the influence of social bias~\cite{hertz_social_2016, chiang} and improving the accuracy of decisions~\cite{collective_intelligence_better_than_independent}. However, depending on the cohesion of the group, the precision of the decision can be overridden by a desire to arbitrarily form a group consensus~\cite{group_think}.

Algorithms have been proposed to increase efficiency and mitigate the influence of subjective external factors found in human decision--making. Many of these systems are powered by black--box AI algorithms and boast high degrees of competency in decision-making tasks across a broad range of contexts~\cite{choi_using_2020, bory_deep_2019, sutton_overview_2020, badue_self-driving_2021}. Although AI can be more efficient than humans, it is also prone to learning biased representations of phenomena through training on historical or socially biased datasets~\cite{mehrabi_survey_2022}. Indeed, AI's ability to discriminate against protected social groups has been found in systems which do not directly use demographic data, due to the presence of indirect proxy relationships~\cite{fairness_through_awarness, assessing_fairness_under_unawareness} -- a phenomenon predicted as far back as 1991 \cite{alan_proxies_causing_human_issues}. Furthermore, despite AI systems typically delivering increased accuracy and efficiency over human decision-makers, they are not always better at making decisions than humans: in their 2018 paper, Dressel and Farid \cite{dressel_accuracy_2018} found that non--expert humans outperformed AI in the task of criminal re--offence prediction, with non--experts predicting more accurately and fairly than AI. 

Taking into account the efficiency and utility benefits of AI, along with the ethical concerns surrounding its inherent lack of transparency and propensity to reinforce systemic biases, research discussions have started to converge towards using AI to enhance decision--making processes through a hybrid human--AI approach. In this manner, humans can act as part of a team alongside AI--assistance, or in a supervisory capacity, for hybrid decision-making.

\subsection{Hybrid Decisions: Human--AI and Human--FairXAI}
One of the main goals of human--AI teaming is to enhance decision-making performance and consistency while maintaining human responsibility. Indeed, early work indicated initial promise in the field, having observed increased accuracy performances of human--AI teams over independent human decisions~\cite{reverberi_experimental_2022}. However, it quickly became apparent that augmenting human performances with AI systems to a level above and beyond what AI systems could achieve by themselves was non--trivial: a plethora of recent work has found that hybrid teams still under-perform compared to baseline AI decisions~\cite{effect_of_confidence_yunfeng, bansal_does_2021, vivian_lai_towards} with `complementary performance' -- a phenomenon whereby human--AI teams exceed the performance of independent human or AI performances -- remaining elusive. Research suggests that this may in part be due to an inability of humans to recognise when AI is providing erroneous assistance, and to counter this, work has begun exploring how explaining AI behaviour may help humans identify erroneous AI--assistance~\cite{effect_of_confidence_yunfeng, bansal_does_2021}. These explanations can come in many forms, from graphical visualisations of features' importance to AI predictive ability \cite{ribeiro2016lime, lundberg2017shap}, through to example-based (counterfactual) explanations of AI decisions \cite{counterfactuals}.

It is thought that, by explaining AI behaviour, human decision--makers will be able to recognise when to trust and ignore AI assistance, resulting in increased performance. However, Zhang et al. \cite{effect_of_confidence_yunfeng} found that XAI techniques (beyond explaining the AI's confidence in its assistance) can have no significant effect on human trust in AI assistance and can make hybrid decisions less accurate. Further research by Bansal et al.~\cite{bansal_does_2021}, found that participants were more willing to accept explained AI assistance but were also no more accurate as a result. Despite the similarity of their experimental setup, both works contrast in their definition of `complementary performance': Bansal et al evaluate complementary performance along a single--dimension, focussing on accuracy in human--AI teams, whilst Zhang et al. provided an insight into the multidimensional aspects of introducing AI explanations on both accuracy and trust. Nevertheless, both papers neglect to investigate how explanations can be used to enhance fairness in human--AI settings, reflecting a broader gap in knowledge as a result of limitations in current empirical work. This is especially important, since recent work has found AI capable of transferring biases to those who interact with it~\cite{humans_inherit_bias}. Furthermore, in light of the accuracy-fairness tradeoff~\cite{accuracy_fairness_tradeoff}, assessing complementary performance in terms of accuracy alone risks driving unfair decision--making practices. Finally, both papers used different XAI techniques across unique decision--domains to explain AI behaviour to human decision--makers, including: graphical visualisations conveying each feature's importance towards AI recommendations whilst predicting income salaries \cite{effect_of_confidence_yunfeng} and a blend between graphical visualisations of AI confidence and text--based explanation \cite{bansal_does_2021}. This makes it difficult to judge the particular effect of the modality of AI explanation on end--user behaviour, but lay open the path for further research.

\subsection{Algorithmic Fairness and XAI}

Researchers have developed multiple techniques to explore bias in AI systems: typically, these involve comparing AI's decision--making performances across different demographic groups, mapping disparities in performance to violations of statistical notions of fairness. These include notions for: group fairness, individual fairness and subgroup fairness ~\cite{mehrabi_survey_2022}. One such fairness definition is `demographic parity' (also known as statistical parity), which conforms to the notion of group fairness and stipulates that all demographic groups should expect similar outcomes~\cite{dwork_fairness_2012}. It was initially anticipated to help bring disadvantaged minority groups to the same degree of representation as majority groups. However, when there exists differences in base rates between groups, demographic parity would be arbitrarily satisfied by selecting unqualified members from a minority group until their acceptance rates are equal to that of the majority group. Overcoming the arbitrary selection criteria of demographic parity, Hardt et al.~\cite{hardt_equality_2016} devised the equality of opportunity fairness metric. Rather than ensuring broad group parity, equality of opportunity enforces that only the selection rates of those who qualify for selection from advantaged and disadvantaged groups, should be equal -- that is, true positive rates should be equal between demographic groups. The issue with equality of opportunity is that it can be satisfied whilst maintaining a representational gap between groups if there is a significant difference in their size. In the same paper, another fairness metric is proposed in the form of equality of odds. Equality of odds stipulates that rates of those who are qualified and selected and non-qualified but selected regardless, must be the same for the advantaged and disadvantaged groups. Whilst these are three of the most commonly referenced statistical definitions within algorithmic fairness literature, many alternatives exist~\cite{chiang, mehrabi_survey_2022}.

Prior work typically focuses on testing AI systems for their ability to conform to fairness definitions \cite{dwork_fairness_2012,hardt_equality_2016}, however, there is a growing interest in deploying tools to promote fairness through human--agency: from enabling users to choose which features are used by AI for fairer inference \cite{Nakao_2022} through to visually inspecting AI models for signs of bias~\cite{yan_iui_rebuttal, cheng_iui_rebuttal} and even visualising the fairness--accuracy tradeoff of an AI model to help designers tailor its performance~\cite{accuracy_fairness_tradeoff}. However, to date, no prior work has ventured to explore the potential of explicitly visualising the fairness--definition derived social bias within an AI model for use as an XAI technique within a deployed decision--making context.

In a separate but growing field of study, researchers have begun to explore the human perception of fairness surrounding the use of algorithmic decision--making~\cite{soliciting_stakeholder_fairness_notions} and machine--behaviour~\cite{van_berkel_information_presentation_impact_on_fairness}, including those featuring accuracy--centric explanations~\cite{dodge_iui_rebuttal}. However, a lack of standardised approaches to eliciting perceptions of fairness makes it difficult to generalise results even between those studies attempting to investigating the same phenomena.

Whilst attention remains focussed on exploring perceptions of fairness of both the use and behaviour of AI in decision--making contexts, little is known regarding how exposing the social biases hidden within AI systems to humans through XAI techniques may affect human--AI decision--making. In the next section, we provide an overview of our methodological approach to assessing how fairness explanations affect decision--making behaviour in hybrid human--AI settings. 

\section{Research Methodology}

\subsection{Overview of Study and its design}

\begin{figure}[h]
    \centering
    \includegraphics[width=0.95\linewidth]{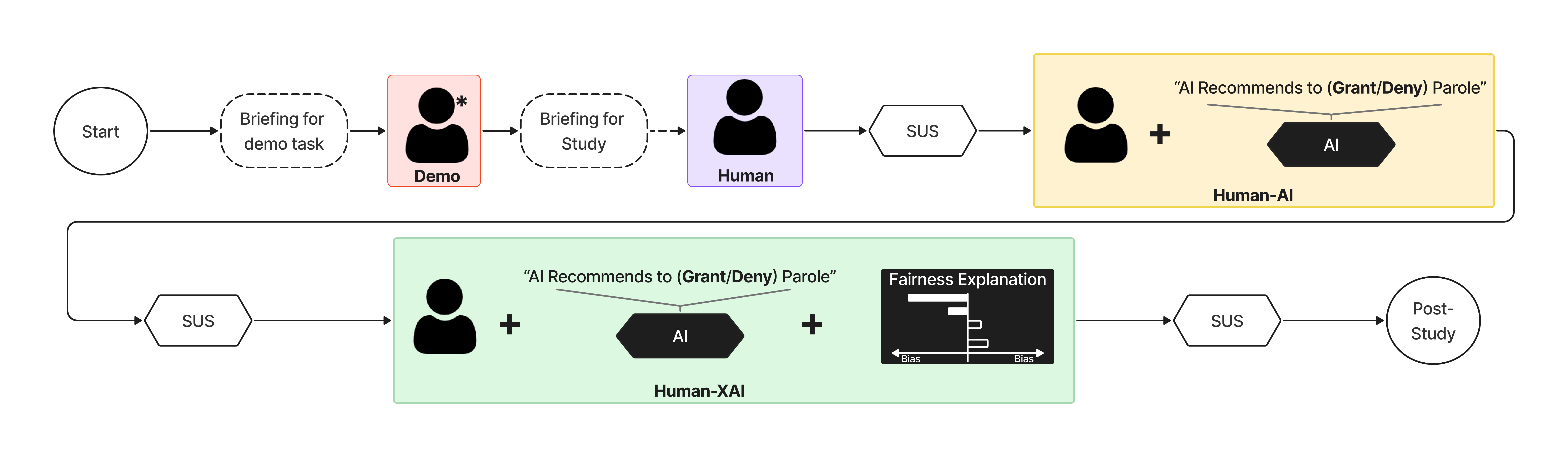}
    \caption{An overview of the experimental procedure. Intervention order was counterbalanced according to participant sex and ethnicity, to mitigate learning-- and order--effects. System Usability Scale (SUS) questionnaires were deployed after each intervention in an attempt to understand how our Fairness Explanations impacted usability of our platform.}\label{fig:simple_research_flow}
\end{figure}

To evaluate the impact of fairness--definition derived AI explanations (henceforth referred to as `Fairness Explanations') on the fairness and utility of hybrid decision-making, we conducted a counterbalanced within--subjects human--participant study ($n = 48$) with recruitment handled through Prolific's \cite{prolific} web platform. The study comprised three interventions that investigated decision-making in unassisted (\textbf{Human}), AI--assisted (\textbf{Human--AI}) and fairness explained AI--assisted (\textbf{Human--FairXAI}) settings. All participants experienced all interventions, and to mitigate undesirable order effects, we counterbalanced the orders in which participants experienced interventions~\cite{greenwald_counterbalancing_within,within_subjects_counter_balanced_web,counter_balanced_cognitive_psychology}. Please see Figure~\ref{fig:simple_research_flow} for an overview of the lab study methodology. Additionally, for a complete breakdown of experimental procedure, please refer to Appendix~\ref{appendix: fair_methodology}, Figure~\ref{appx:study_methodology}. This design enabled us to explore the ability of fairness explanations to induce fairer behaviour over traditional unassisted (\textbf{Human}) and AI--assissted (\textbf{Human--AI}) decision-making settings, by directly controlling for order and learning confounds.

The scenario used in this study was based on the criminal recidivism domain explored in previous work~\cite{dressel_accuracy_2018,human_limits_prediction_no_immediate_accuracy, chiang}, adapted to include parole decision making to situate it within a real--world high--stakes setting. This adaptation was necessary to observe the effects of fairness explanations on decision--making as opposed to predictions -- predictions and decisions differ in nature~\cite{differences_between_predictions_decisions_finance_markets}. Furthermore, judges are reported to use re--offence risk predictions in parole decision--making scenarios \cite{predicting_risk_in_criminal_procedure}, 
thus the adaptation of the domain served to increase the ecological validity of the previously AI--centered prediction task.

In the \textbf{Human} treatment, participants were shown parole-candidate profiles sampled from a fairly balanced synthetic dataset (see Section~\ref{datagen}) and asked to grant or deny parole. Throughout the \textbf{Human--AI} treatment, participants worked alongside an AI--recommendation system for the same task. Recommendations were generated in a systematically balanced manner with respect to ground truth and were delivered in the form of text prompts (i.e. ``AI recommends to grant parole'' or ``AI recommends to deny parole''). In the \textbf{Human--FairXAI} treatment, AI-recommendations were paired with fairness explanations in a systematically balanced manner using an extension of the strategy deployed in the \textbf{Human--AI} treatment. This meant that participants experienced working alongside an equal number of models presented as containing low-- and high--bias. Fairness explanations were simulated using a random number generator and were stratified into low-- and high--bias explanations depending on whether the explanations' absolute bias quantification exceeded a threshold -- initialised from prior literature~\cite{people_see_more_bias_in_algorithms} as 20\% and further tuned with feedback from piloting.

The following three subsections describe our research methodology in depth, beginning with how we generated the data used within the study, followed by an explanation of our novel approach to designing and generating contextualised fairness explanations from user--feedback. Finally, we elaborate on the experimental task performed by participants in our online study.

\subsection{Data Generation}
\label{datagen}

A dataset commonly used to understand fairness in decision making was compiled by ProPublica and is based on risk scores generated by Northpointe's (now Equivant) COMPAS system \cite{compas_dataset}. COMPAS has been described as being used as an aid in the US judicial system for pre--trial, probation and parole decision--making, by informing judges of the AI--estimated risk of defendants re--offending. COMPAS provides a set of data pertaining to historical crimes committed in Broward County, US, and includes: recidivism risk score, the ground truth label for whether or not an arrested individual re--offended, in addition to data on their criminal histories and demographic information such as their sex and race. In 2016, COMPAS was reported to propagate racial biases when profiling arrested individuals~\cite{angwin_machine_2016}. Thus, to observe the fairness and utility of participants' decisions (\textbf{RQ1}) without them being influenced by the systematic biases embedded within the COMPAS data, we needed to ensure that defendant profiles were sampled in a fair manner. 

\begin{figure}[h]
    \centering
    \includegraphics[scale=0.5]{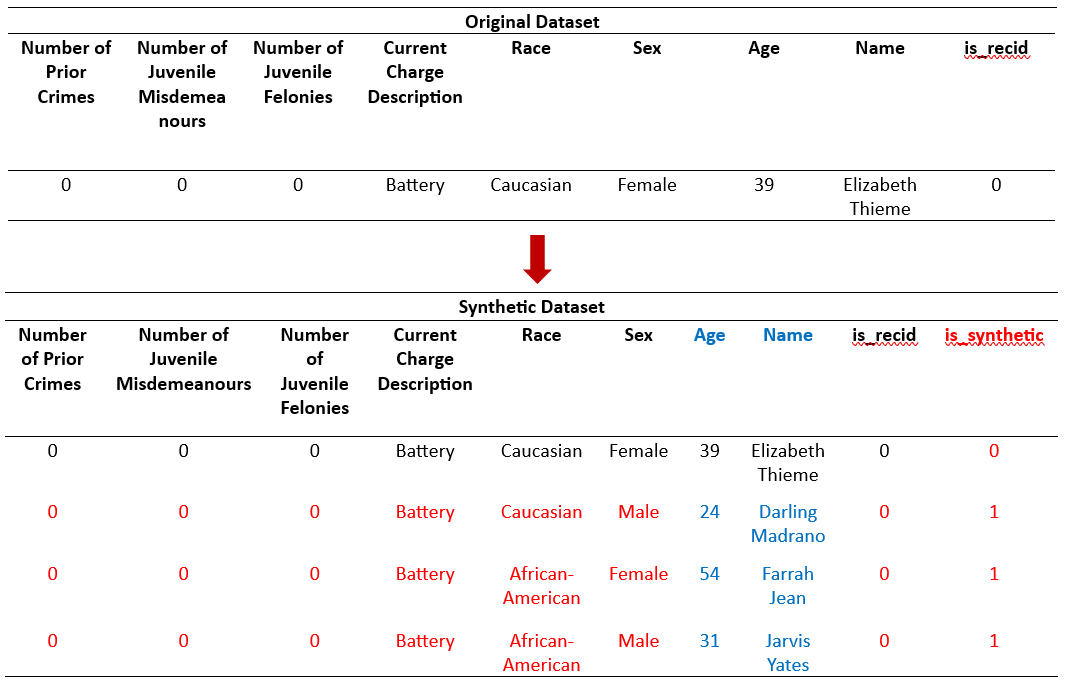}
    \caption{
    We create new synthetic instances of each criminal history and apply them equally across protected characteristics (denoted with {\color{red} red text}). To remove proxy associations between criminal histories and secondary characteristics (Name, Age), we take a random sample from the sub-populations (Name $\times$ Sex $\times$ Race) and pair them with the new synthetic instances (denoted with {\color{blue} blue text}).}
    \label{appx:pc_flipping}
    \Description{A row of original COMPAS data containing criminal profiles is selected and three new synthetic instances of that row are created with all possible Race and Sex group combinations. These new Race and Sex groups are attached to the synthetic rows, and new name and age information is generated for them from the original distributions. The synthetic rows are labelled with the `is_synthetic' field set to `1', whilst the original row retains a value of `0' to signify it is original.}
\end{figure}

Previous work~\cite{dressel_accuracy_2018, chiang} are reported to have tackled this challenge by randomly sampling the COMPAS dataset. Unfortunately, we were unable to obtain the datasets used in~\cite{dressel_accuracy_2018, chiang} thus it is unclear whether the data they used for their studies reflected the biases held in the original COMPAS dataset. Nevertheless, due to the non--uniform distribution of data in the original dataset, randomised sampling from the union of the sets of Caucasian and African--American defendants would yield an over--representation of African--American samples (primarily recidivist) over all other groups. As a result, decision--makers could form unfair negative perceptions of sample populations as a result of sampling strategy alone. To mitigate over--representation bias in the original dataset, researchers may rather opt to sub--sample according to protected characteristic categories (i.e., sex and race). However, even this approach does not explicitly dissociate the defendants' protected characteristics from their criminal histories and profiles, meaning that any embedded relationships between race, sex and criminal histories will remain as a potential bias in the sample conveyed to participants. As a result, statistical stereotypes are likely to be maintained through this approach and could serve as a confounding bias in evaluating the fairness of decision--makers. 

To mitigate sampling bias and stereotypes in our sampled data, we extended current techniques to synthetically equalise base--rates between protected groups. We began by dissociating protected characteristics ($\in${ [\textit{race, sex}]}) from criminal histories ($\in$ [\textit{number of prior crimes, number of juvenile misdemeanours, number of juvenile felonies, charge classification, current charge description}]) by randomly sub--sampling each protected group uniformly, with respect to the outcome label -- the eventual re-offence status of the defendant ($\in$ [\textit{is\_recid}]). Once we had uniform, random sub--samples of the COMPAS dataset according to the product of sex, race and ground-truth ($\in$ [\textit{sex $\times$ race $\times$ is\_recid}]), we dissociated protected characteristics and criminal histories through the creation of a synthetic dataset with equalised base rates by using the protected characteristic flipping method~\cite{protected_characteristic_flipping}. This method generates a new synthetic instance of data for each alternative protected characteristic grouping of the original criminal history data and the same ground truth value. With the combination of uniform random sampling and protected characteristic flipping, we eliminate the risks of our data being confounded by sampling biases and stereotypes embedded in the original historical data.

\begin{figure}[h]
    \centering
    \includegraphics[scale=0.6]{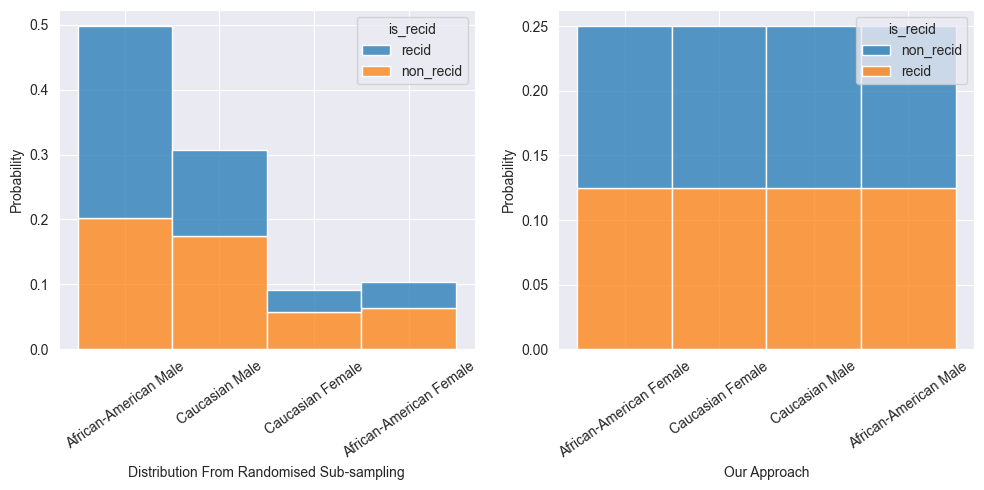}
    \caption{The graph on the left shows the distribution of a 1,000 random sample taken from the COMPAS dataset, whilst the graph on the right shows the distribution using our method which includes the protected characteristic flipping method \cite{protected_characteristic_flipping}}.
    \label{fig:randomised_vs_ours}
    \Description{The left graph shows a probability distribution plot (generated using the sampling methodologies in \cite{chiang, dressel_accuracy_2018}) of Race and Sex groups against their probability of re-offending and not re-offending. The data shows a skew towards African-American Males with a high probability of comitting re-offence compared to caucasian males. On the right is a second graph which shows the probability distribution plot generated via our sub-sampling + protected characteristic flipping method that retains an equal 50/50 split in all race and sex groups probability of re-offending and not re-offending.}
\end{figure}

Once protected characteristics and criminal histories were dissociated, we paired the synthetic data with profile pictures to increase realism during decision--making and provoke an increase in observable unconscious bias exhibited by individuals during our study. Suitable profile pictures were extracted from the FairFace \cite{karkkainenfairface} dataset: a public dataset of images of human faces taken in the wild and labelled according to their protected characteristic groups: age, sex, and ethnicity. Since FairFace is comprised of in--the--wild pictures, there are often artefact, object, and lighting discrepancies between photos that make it hard to clearly distinguish the human subject of the photo. To mitigate the risk of these anomalous photos confusing our participants, we used OpenCV's \cite{opencv_library} frontal face detection tools to ensure that each photo features a face visible to the human eye -- an example of these images can be seen in Figure \ref{fig:human-task}. We then pooled these filtered images according to their Age, Sex and Ethnicity labels and randomly paired them with their respective features within defendant profile data. Images were removed from the pool on a per-participant basis, meaning that each participant would never see the same image twice. Additionally, as the photos from FairFace are scraped from the internet and feature depictions of real--persons, we included a disclaimer within the introduction sequence of our online study that informed participants of the nature of the images and clarified they were included as a visual aid and were intended to be consumed for reference purposes only. As a contribution to the research community, this `fair'--COMPAS dataset will be made available via open access from GitHub.

\subsection{Iterative Design for Visual Fairness Explanations and Algorithmic Approach to Deriving Fairness Scores}
\label{fpi}

\begin{figure}
    \centering
    \includegraphics[width=0.95\linewidth]{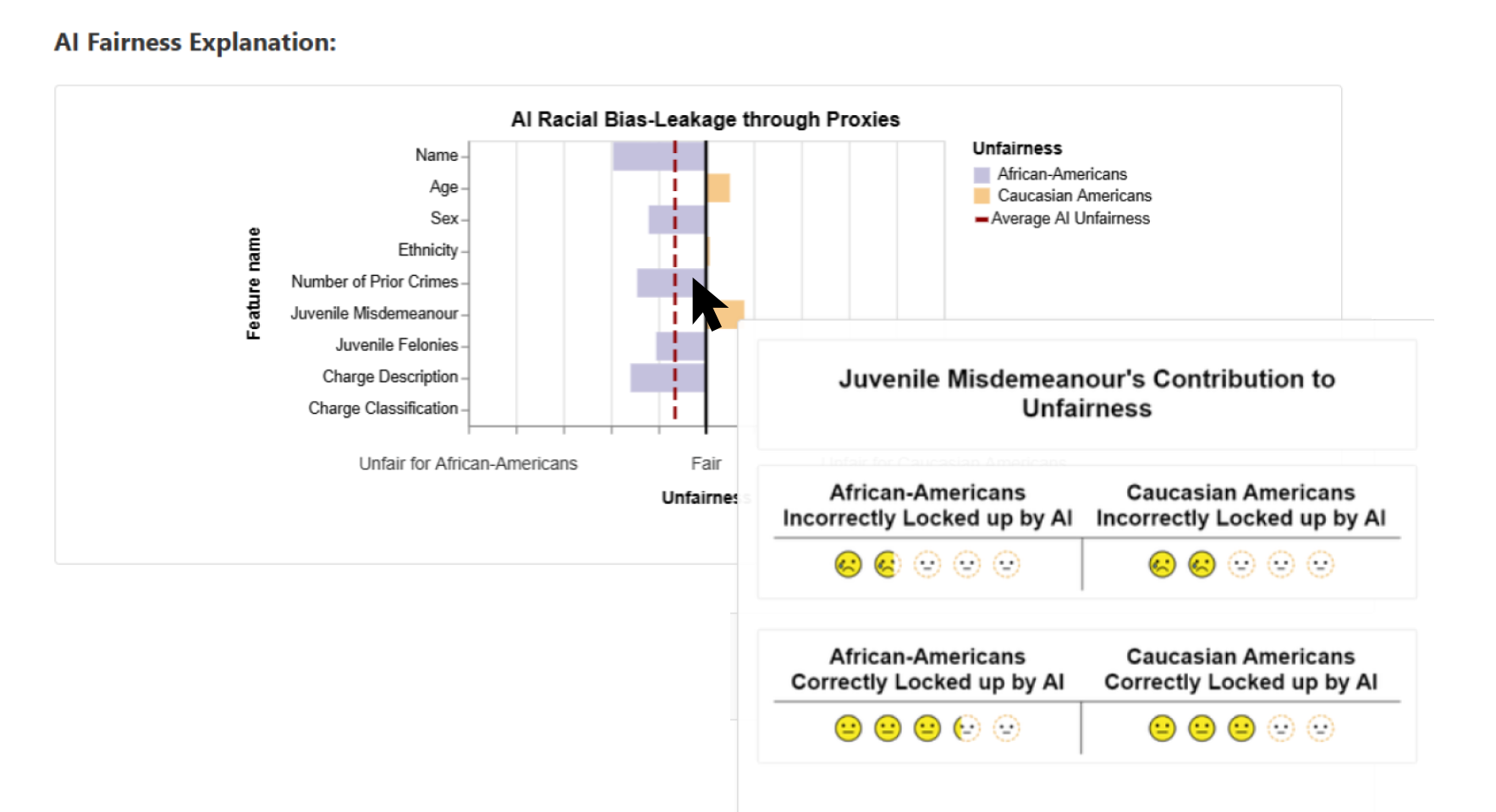}
    \caption{This figure shows the final fairness explanation design after performing iterative design with pilot participants. The explanations consisted of three main features: magnitude and direction of the AI's overall bias (vertical red  dotted line), contribution of each feature towards global bias (horizontal bars) and downstream impact of bias caused by each feature (tooltip)}
    \label{fig:fairness-explanation-tooltip}
\end{figure}

Our main research goal was to establish the impact of incorporating fairness explanations into hybrid human--AI decision-making settings at decision time (\textbf{RQ1}). To explore this concept, we needed to provide participants with a means of exploring bias in AI models. Inspired by related work in visual--XAI~\cite{lundberg2017shap,ribeiro2016lime}, we performed a series of iterative discussions with pilot participants from a range of engineering--based disciplines, aimed at adapting pre--exisiting, accuracy--centric visual explanations to help facilitate the identification of social bias in AI models. These discussions identified key features perceived to be useful in identifying bias in AI models, namely: quantity of bias and direction of bias. Further participant led discussions resulted in the ideation of an additional interactive tooltip to accompany the fairness explanation, to convey the downstream impact of unfairness caused by each feature. The final design can be seen in Figure~\ref{fig:fairness-explanation-tooltip}. 

The main features of the explanation include the: 

\begin{itemize}
    \item global bias of the AI model (denoted by the x--axis position of the red vertical dashed line) derived from the selected fairness definition,
    \item contribution of each feature towards global bias (horizontal bars),
    \item downstream impact of the bias caused by each feature (tooltip) to aid comprehension of the real--world impact of the AI's bias.
    
\end{itemize}

To ensure participants in our empirical study (section~\ref{subsec:experiment_design}) were likely to comprehend the information conveyed through our fairness explanations, we first piloted a series of pre--generated explanations with a handful of participants who had no prior experience interpreting them. Participants reported satisfactory comprehension of two of the major features of the graph (direction and magnitude of global bias of the AI model and each feature's contribution towards that bias), but struggled internalising the downstream bias conveyed through the tooltip. In an effort to overcome these challenges, increase comprehension by non--technical experts and thus reduce noise, we created a short introductory video which explained the three main features of the explanation chart. This video appeared directly before answering the first question in the \textbf{Human--FairXAI} treatment condition.

Since we did not have access to the AI model used by COMPAS to generate risk scores, we generated synthetic explanations using random--number generators and paired these synthetic explanations in a balanced manner with respect to AI advice However, in the next section (Section~\ref{section: FPI}), we describe a potential post--hoc technique that could be used to generate visual fairness explanations, using a variation on the popular Permutation Feature Importance algorithm~\cite{PI}.

\subsection{Fair Permutation Importance}
\label{section: FPI}
To assess the importance of each feature in any given AI prediction model, we must first derive the permutation feature importance for that feature. Fairness scores depend on the definition of fairness used, for example: equality of opportunity and equalized odds \cite{hardt_equality_2016} are calculated as the percentage difference between the model's ability to satisfy fairness criteria between the advantaged, and disadvantaged group. To calculate a fairness score $S$ of a model for equality of opportunity ($S_{EoP}$), one can perform the following calculation: \[S_{EoP} = (TPR_{majority} - TPR_{minority})\] where $TPR$ denotes the True Positive Rate of the majority or minority group. For equality of odds, we adapt the fairness score equation to include the True Negative Rates $TNR$ of the majority and minority groups: \[S_{EoD} = (TPR_{majority} + TNR_{majority}) - (TPR_{minority} + TNR_{minority})\]

For the purposes of the explanations used in our study, we wanted to use a consistent framing \cite{positive_framing} to improve understanding in our contextualised fairness explanations. Therefore, we opted to show unfair discrepancies in the Positive Predictive Value (PPV) performance of AI for protected groups, to our participants. The PPV fairness score can be calculated as follows: \[S_{PPV} = (\frac{TP}{TP + FP})_{majority} - (\frac{TP}{TP+FP})_{minority}\]

\begin{landscape}
\begin{figure}
    \centering
    \includegraphics[width=\linewidth, height=\textheight, keepaspectratio]{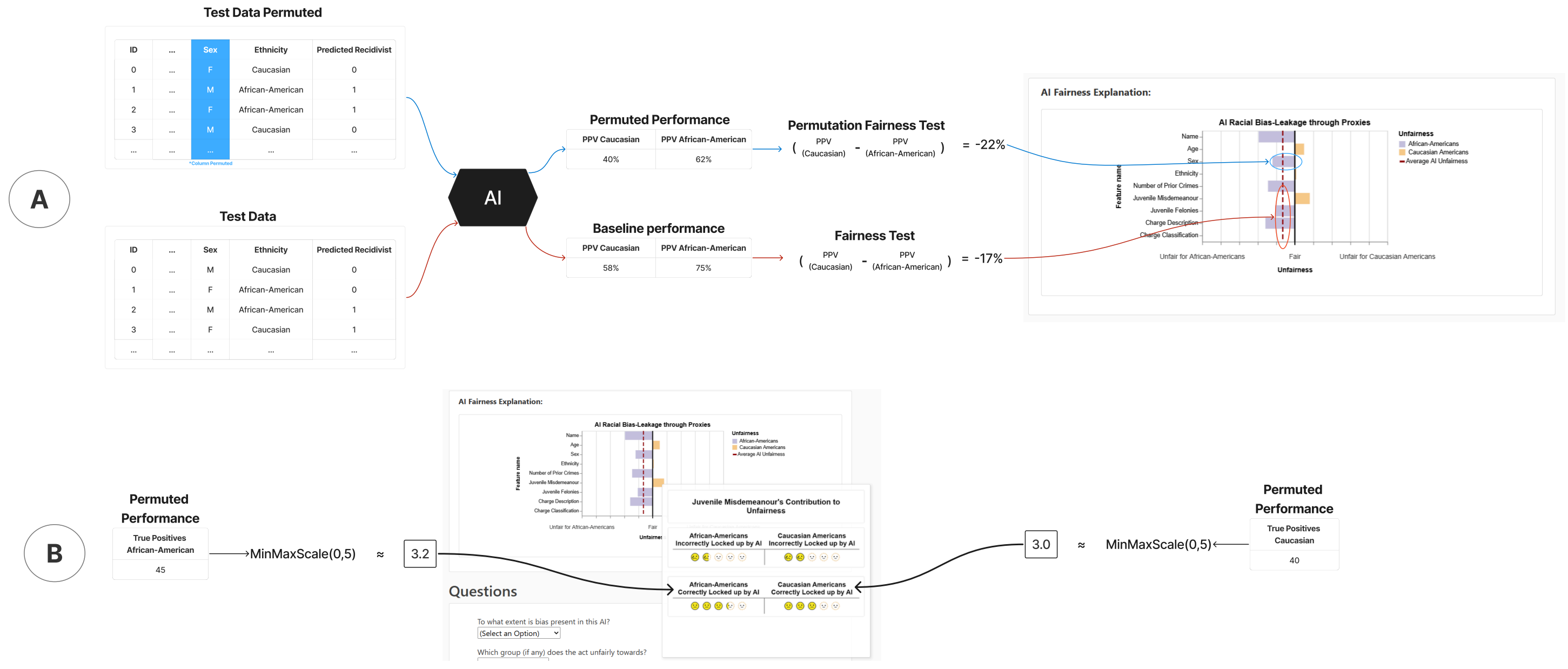}
    \caption{A represents how the FPI mechanism may be used to generate feature contributions towards the model's unfairness, whilst B represents how permuted fairness scores can be rendered using a tool--tip to contextualise the downstream impact of the bias caused by that feature.}
    \label{fig:fpi_mechanism_detail}
\end{figure}
\end{landscape}

Once the fairness scores have been calculated for a given model, we can identify the impact and direction of bias that each feature contributes towards AI unfairness by randomly permuting the data on a feature-by-feature basis. The importance of fairness for each feature $i_{j}$ can be calculated as follows: \[i_{j} = S - \frac{1}{K}\sum_{k=1}^{K}S_{kj}\] where $j$ denotes the feature permuted, $S$ is the fairness score of the model and $K$ is the number of permutations. After permuting the data and arriving at the fairness importance $i_{j}$, we can use the mean and variance of the contribution to demonstrate that the feature contributes to unfairness. To generate the tooltip observed in Figure~\ref{fig:fairness-explanation-tooltip}, we take the normalised ratio of fairness criteria used to generate the fairness scores for the AI model. Please refer to Figure~\ref{fig:fpi_mechanism_detail} for a visual representation of how we use this approach to generate data suitable for use by the designs proposed in section~\ref{fpi}.

\subsection{Experimental Design}
\label{subsec:experiment_design}
To test the ability of fairness explanations to induce fairer behaviour in non--expert decision-makers, we devised a counterbalanced, within--subjects lab experiment that establishes decision-making behaviour with and without AI-assistance.

\subsubsection{Ethics Statement}
The research team did not identify any risks to the health or well-being of the participants in this experiment. Our institutional ethics review committee granted ethics approval prior to the study taking place and consent with the right to withdraw was obtained from participants prior to participating in the study.

\subsubsection{Demographics}

Participant recruitment was conducted through Prolific's \cite{prolific} online platform and enrolment was open for four days. This experiment was single--phase, with participants expected to complete it within the same day as enrolling. Following ethical guidelines~\cite{payment_enough}, participants were paid at an hourly rate of £11.54. We recorded an average completion time of 60 minutes with a standard deviation of 23 minutes ($M = 60.31, SD = 23.508$). Recruitment conditions limited participants to only those living and born in the US, since pilot studies informed us that terminologies used within COMPAS's criminal profiles (specifically, crime descriptions) were not easily interpreted by a cross-cultural English-speaking audience (e.g., US--UK). We recruited 48 participants $(n = 48)$, comprising: 12 African--American male, 12 African--American female, 12 Caucasian male and 12 Caucasian females. The average age of participants was 37 with a deviation of 13 years  ($M = 37.23$ $SD = 13.40$).

\subsubsection{Experimental Task \& Treatments}
For this experiment, we used the context of recidivism prediction and parole sentencing. The scenario is based on previous experiments performed by Dressel and Farid (2018) ~\cite{dressel_accuracy_2018} and Chiang (et al. 2023)~\cite{chiang}, albeit adapted to facilitate decision--making. Participants were shown profiles of ``parole candidates'' -- convicted criminals, eligible for parole -- in a random--order, before being asked to decide to grant or deny parole. All parole candidates had the same features as previous experiments -- that is: full name, age, sex, ethnicity, number of prior crimes, number of juvenile felonies, number of juvenile misdemeanours, current charge category and current charge description -- and these variables served as the basis from which participants made parole decisions across all interventions. \\

\begin{itemize}
    \item \noindent\emph{Unassisted (Human) treatment:}
In the unassisted treatment, participants made decisions using basic parole candidate information. Participants made \textbf{8} decisions, representing balanced coverage across identified protected characteristics (sex and ethnicity), in relation to the outcome variable (is\_recid). Formally: \textbf{Human} $=$  \textbf{sex} ($\in$ [male, female]), \textbf{ethnicity} ($\in$ [african--american, caucasian]), \textbf{is\_recid} ($\in$ [recidivist, non--recidivist]).\\

\item \noindent\emph{AI--assisted (Human--AI) treatment:}
 In the AI--assisted treatment, participants received an additional variable that listed the AI's recommendation as either: `AI recommends to Grant Parole' or `AI recommends to Deny Parole'. These recommendations were paired in a systematically balanced manner with respect to the protected characteristics and ground truth. As a result, participants answered \textbf{16} decision--tasks, to account for the additional feature: Human--AI = \textbf{Human} (sex, ethnicity, is\_recid) + \textbf{AI recommendation} ($\in$ ["AI recommends to Grant Parole", "AI recommends to Deny Parole"]).\\

 \item \noindent\emph{Fairness explained, AI--assisted (Human--FairXAI) treatment:}
 In the fairness explained treatment, participants were provided with an accompanying visual fairness explanation to enable them to observe the AI's global bias (high-- and low--bias). These fairness explanations were based on the design shown in Figure~\ref{fig:fairness-explanation-tooltip}, but the data was simulated using a pseudo-random number generator and stratified according to a threshold for low-- and high--bias and counterbalanced in relation to AI--recommendations. As a result, participants answered \textbf{32} questions in the Human--FairXAI treatment, to account for the introduction of another variable: \textbf{Human--FairXAI} $=$ \textbf{Human--AI} $+$ \textbf{XAI} (low--, high--bias).
 
\end{itemize}

\begin{figure}[h]
    \centering
    \includegraphics[width=0.5\linewidth]{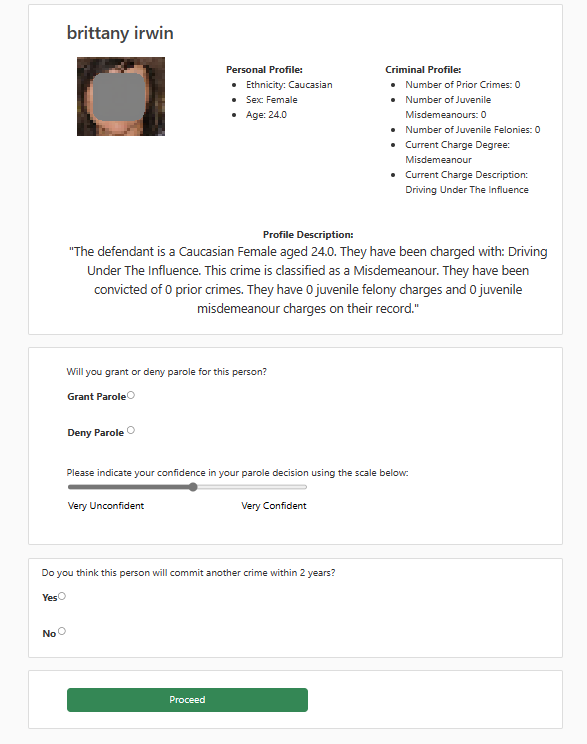}
    \caption{A typical example of a case a participant might encounter in the \textbf{Human} decision-making condition. Participants were expected to provide a decision given the data presented to them. The image of the face of the parole candidate has been anonymised for the purposes of publication and dissemination, since candidate profile images were originally extracted from a public dataset. Any likenesses of the face data to real-world individuals does not imply real-world criminal activity nor intent - a fact to which participants in our study were made aware of before beginning the study. Suitable generative AI tools that could produce lifelike synthetic representations of a human facial profile were not commonly available at the time of conducting this study.}
    \label{fig:human-task}
\end{figure}

\noindent\emph{Question Structure:}
Across all treatments, participants were instructed to take into account all available information before making their final parole decision. For decision--making, participants were asked the question \textit{``Will you grant or deny parole?''} and could choose a response from either `grant parole' or `deny parole'. To predict recidivism, participants responded to the same question prompts used in earlier studies -- \textit{``Do you think this person will commit another crime within 2 years?''} -- to which they responded either `Yes' or `No'. Finally, participants were asked to provide a self--assessment of their confidence in their decision, to establish whether the introduction of fairness explanations had any impact on the confidence in combined human--AI decision--making. For an example of the task performed by participants, refer to Figure~\ref{fig:human-task}.\\

\noindent \emph{Design Rationale:}
To ensure inherent variability in our experimental design due to an increasing number of questions per treatment condition (8, 16, 32) was not confounding our results, we performed a monte--carlo simulation using synthetic data prior to running our empirical study. We found the variability introduced by the design of our study to be non--significant, which increased conviction in our experimental design.

We opted for simulating both AI--recommendations and explanations to retain control over the magnitude and direction of bias each model's recommendation was conveying to participants. Since we could not run post-hoc evaluation techniques on COMPAS' prediction system and we could not guarantee that AI recommendations from a bespoke model were themselves free from bias or that their biases were exhaustively accounted for, the simulation of AI and explanations was necessary to ensure we retained control over the experimental protocol. We recognise this limits the generalisability of our study's findings towards real--world performant AI (in particular with regards to its ability to persuade), but nevertheless this work serves as an important empirical stimulus to support future research to explore how AI performance impacts the effectiveness of fairness explanations in reducing bias in combined decision--making.

Within each intervention, participant attention was checked through mandatory checks. Attention checks were presented as normal questions, however, the parole candidate information was replaced by directives which instructed participants to respond in a particular (randomised) manner. All participants passed all checks, however, would have been excluded from analysis if they failed to answer the majority of attention--checks correctly. Additionally, between interventions, participants were asked to perform a System Usability Scale (SUS) \cite{sus_scale} evaluation of that intervention's usability. This was performed in an effort to understand how fairness explanations might impact the usability of human--AI platforms and to explore potential trade--offs between usability and performance.

In the next section, we describe the results from our online study and the tests performed as supporting evidence for fairness explanations' roles in relation to key research questions.

\section{Results}
\label{sec:results}
We used the \textit{is\_recid} label in the COMPAS dataset~\cite{compas_dataset} to act as a ground truth for our parole decisions, as it denotes the eventual recidivism (if any) of each defendant. For the sake of consistency with prior work wherein the positive condition indicates criminal re-offence ~\cite{chiang, compas_dataset}, we chose the \textbf{deny parole} response to correspond to a positive prediction and the \textbf{grant parole} response thus corresponds to a negative prediction. As a result, a criminal who would eventually go on to commit a future crime and who was denied parole was classified as a true positive and a non-re-offending criminal who was granted parole was classified as a true negative.

\subsection{Accuracy}
\label{sub:results_acc}

To assess the influence of fairness explanations on accuracy in decision-making (\textbf{RQ1.a}), we performed statistical analysis on two measures of participant accuracy: Accuracy and PPV.

We tested accuracy across treatment arms -- \textbf{Human} ($M = 60.16, SD = 15.82$), \textbf{Human--AI} ($M = 59.38, SD = 11.60$), \textbf{Human--FairXAI} ($M = 59.81, SD = 9.63$) -- using a repeated measures ANOVA at a 95\% confidence interval.  After correcting for multiple tests, we found no significant difference in accuracy scores across treatment conditions: $F$(1.563, 73.456) = 0.066, $p$ > 0.05 and a negligible effect $\eta^2$ = 0.001.

When testing for accuracy performances of participants according to PPV (the metric used within the fairness explanation itself), a repeated measures ANOVA found a significant difference between at least two of the treatment arms:  $F$(2, 94) = 12.407, $p$ < 0.01 with a large effect size $\eta^2$ = 0.21. Post-hoc evaluations indicated that the \textbf{Human--FairXAI} ($M = 62.76, SD = 16.25$) condition significantly improved participant performance over both the \textbf{Human} ($M = 45.31, SD = 19.743$) treatment ($p$ < 0.01, 95\% C.I. = [-27.10, -7.79]) and the \textbf{Human--AI} ($M = 53.472, SD = 20.400$) treatment ($p$ = 0.029, 95\% C.I. = [-17.81, -0.76]). In addition, the \textbf{Human--AI} treatment was found to significantly improve participant PPV over \textbf{Human} ($p$ = 0.038, 95\% C.I. = [-15.98, -0.34]).  Since PPV measures the accuracy of positive predictions made by participants, we find that fairness explanations helped participants make more accurate \textbf{deny parole} decisions compared to AI-assisted and unassisted conditions.

Please see Figure~\ref{fig:mean-accuracy}, for a visual representation of accuracy means and Figure~\ref{fig:mean-ppv} in the appendix for a visual representation of PPV means across experimental conditions.

\subsection{Fairness}
\label{sub:fairness}
To evaluate the ability of fairness explanations to improve fairness (\textbf{RQ1.a}), we compared fairness delta metrics \cite{chiang, ppv_fairness_metrics} for each protected group category (sex, ethnicity). Concretely, absolute deltas were derived on a per--participant basis from the difference between their performance metrics (i.e., accuracy) for parole candidates' associated protected groups (i.e., male and female).  Non-zero deltas represent biased decision--making behaviour. The fairness metrics that we tested were: $\Delta$Accuracy, $\Delta$PPV, $\Delta$POS, $\Delta$FPR, $\Delta$FNR, consistent with prior work~\cite{chiang}. We did not compare the $\Delta$Twin metric found in~\cite{chiang}, as our methodology did not enable us to explore participant decisions on direct twins. In addition, our data differs from prior work~\cite{chiang}, in that we measure absolute differences in performance rather than relative differences. We made this change after pilot studies informed us that it was necessary in order to observe net reductions in bias across uniformly sampled demographic groups, wherein equal and opposite biases can cancel out - similar in effect to the phenomenon seen in decisions made by groups of individuals~\cite{chiang}. Finally, we controlled for errors introduced through multiple comparisons by using strict Holm-Bonferroni~\cite{Holm_Bonferroni,holm_bonferroni_agents,holm_bonferroni_new_test} corrections and included the tests in Section~\ref{sub:results_acc} as additional tests. For post--hoc comparisons, as our data did not meet assumptions of equal variance according to levene's test, we performed strict bonferroni corrections. As a result, our post--hoc analyses of statistical tests are likely to have reduced power, but maintain robustness to noise.

\begin{table}[]
\centering
\resizebox{\columnwidth}{!}{%
\begin{tabular}{clllc}
\hline
\multicolumn{1}{c|}{\multirow{2}{*}{\textbf{Metric}}} & \multicolumn{3}{c|}{\textbf{Treatment Arms}} & \multirow{2}{*}{\textbf{$p$-value}} \\
\multicolumn{1}{c|}{} & \multicolumn{1}{c|}{\textbf{Human}} & \multicolumn{1}{c|}{\textbf{Human-AI}} & \multicolumn{1}{c|}{\textbf{Human-FairXAI}} &  \\ \hline
$\Delta$Accuracy (Sex) & $M$ = 24.48, $SD$ = 23.34 & $M$ = 17.19, $SD$ = 9.44 & $M$ = 11.02, $SD$ = 9.44 & 0.00521* \\
$\Delta$Accuracy (Race) & $M$ = 26.56, $SD$ = 22.13 & $M$ = 16.67, $SD$ = 12.19 & $M$ = 11.11, $SD$ = 9.79 & 0.000197* \\
$\Delta$PPV (Sex) & $M$ = 34.38, $SD$ = 31.21 & $M$ = 28.47, $SD$ = 21.60 & $M$ = 20.54, $SD$ = 16.99 & 0.0314 \\
$\Delta$PPV (Race) & $M$ = 35.42, $SD$ = 31.72 & $M$ = 27.78, $SD$ = 18.78 & $M$ = 20.63, $SD$ = 16.97 & 0.0139 \\
$\Delta$POS (Sex) & $M$ = 22.92, $SD$ = 19.18 & $M$ = 18.75, $SD$ = 15.25 & $M$ = 11.66 , $SD$ = 9.44 & 0.00232* \\
$\Delta$POS (Race) & $M$ = 27.08, $SD$ = 22.97 & $M$ = 18.75, $SD$ = 18.05 & $M$ = 14.38, $SD$ = 16.30 & 0.00289* \\
$\Delta$FPR (Sex) & $M$ = 29.17, $SD$ = 30.65 & $M$ = 22.92, $SD$ = 21.78 & $M$ = 12.86, $SD$ = 12.67 & 0.00525* \\
$\Delta$FPR (Race) & $M$ = 39.58, $SD$ = 35.67 & $M$ = 19.79, $SD$ = 23.61 & $M$ = 17.03, $SD$ = 19.47 & 0.000313* \\
$\Delta$FNR (Sex) & $M$ = 36.56, $SD$ = 35.34 & $M$ = 27.08, $SD$ = 22.97 & $M$ = 18.23, $SD$ = 14.58 & 0.00662* \\
$\Delta$FNR (Race) & $M$ = 34.38, $SD$ = 34.45 & $M$ = 26.04, $SD$ = 24.17 & $M$ = 19.79, $SD$ = 18.02 & 0.0329 \\ \cline{2-5} 
\end{tabular}%
}
\caption{Once adjusted for multiple tests, we find that fairness explanations were responsible for significantly reducing deltas in Accuracy, POS and FPR from both a race and sex perspective. Significance was also found for FNR fairness delta when examining sex bias, however, no significance was found when examining race bias. Since the data used to assess fairness is derived from the accuracy performances of participants, we used the Holm-Bonferroni correction over 12 tests, to ensure that the error-rate did not exceed 5\%.}
\vspace{-2em}

\label{tab:fairness}
\end{table}

Table~\ref{tab:fairness} lists all the fairness tests performed: we find that the \textbf{Human--FairXAI} treatment is responsible for significantly reducing bias according to nearly every metric assessed, with the exception being $\Delta$PPV and $\Delta$FNR(Race). This means, fairness explanations were found to significantly improve: accuracy equality ($\Delta$Accuracy ~\cite{accuracy_equality, chiang}), demographic parity ($\Delta$POS~\cite{dwork_fairness_2012, chiang}) and equality of opportunity fairness ($\Delta$FNR~\cite{hardt_equality_2016}).

\subsection{Confidence}
\label{sub:confidence}

To establish how AI decision-support and Fairness Explanations influenced participant confidence in decisions (\textbf{RQ1.a}), we performed a repeated measures ANOVA on the confidence values of grant and deny parole decisions for participants across conditions.

Although no significance was found when testing with the repeated measures ANOVA, confidence scores were lowest for the \textbf{Human-AI} condition ($M = 16.36, SD = 2.42$), with the \textbf{Human} condition scoring the highest confidence ($M = 16.60, SD = 2.34$)  and \textbf{Human-FairXAI} ($M = 16.46, SD = 2.47$) scoring in the middle. See Appendix~\ref{appx:confidence} (Figure~\ref{fig:mean_confidence}) for a graphical representation. This data appears to contradict a phenomenon in previous studies in which end-users tend to have the higher confidence when using unexplained AI-assistance \cite{effect_of_confidence_yunfeng, bansal_does_2021}.

\subsection{Automation Bias}
\label{sub:automation-bias}

To understand the ability of Fairness explanations to reduce automation bias (\textbf{RQ1.b}), we performed a paired t-test on the percentage agreement of participants with AI decisions (regardless of AI correctness) across \textbf{Human-AI} and \textbf{Human-FairXAI} interventions, similar to \cite{automation_bias_measure}. We observed a slight decrease in the reliance on AI, but did not find significant differences between \textbf{Human-AI} ($M = 58.85, SD = 15.60$) and \textbf{Human-FairXAI} ($M = 56.61, SD = 12.12$) conditions when tested with a paired samples t-test: $t(47) = 1.038, p = 0.305$. See appendix~\ref{appx:automation_bias} (Figure~\ref{fig:automation_bias_mean}) for a visual representation of the data.

\subsection{System Usability and Participant Understanding of Explanations}

Finally, we evaluated the usability of the parole decision-making tool to understand how fairness explanations contribute to usable \textbf{Human-FairXAI} systems (\textbf{RQ1.c}). We capture usability factors according to the standardised System Usability Scale (SUS) \cite{sus_scale} and calculate usability scores according to their positive and negative framing \cite{sus_scale_calculation, sus_scale}. The data suggested that the \textbf{Human} treatment was the most usable ($M = 75.80, SD = 16.37$), with \textbf{Human-AI} scoring second highest ($M = 72.81, SD = 15.37$) and \textbf{Human-FairXAI} having the lowest usability score ($M = 65.21, SD = 20.69$). We tested for significance between the conditions using a repeated measures ANOVA and found significance: F(2, 92) = 6.847, p = 0.00190, with a large effect ($\eta^2$ = 0.21). Furthermore, post hoc comparisons found significant differences between the \textbf{Human} and \textbf{Human-FairXAI} treatments (p < 0.01, 95\% C.I. = [4.66, 16.28]). In addition, a significant reduction in SUS scores was found between the Human-AI and Human-FairXAI treatments (p = 0.01, 95\% C.I. = [1.55, 13.66]). Thus we conclude that introducing visual fairness explanations holds the potential to significantly reduce the usability of a system.

Regarding participant understanding of the fairness explanations at the time of decision--making: we asked two questions pertaining to the high--level features of the explanations to ensure participants had a basic understanding of the information presented to them and to ensure that the effects we observed were not the result of some unconscious faculty. Concretely, participants were asked to identify the quantity of bias in the AI's recommendation and to which group (if any) that bias was causing unfairness towards. We found participants exhibited a high--degree of understanding of the basic information conveyed through the fairness explanations (an average of 65\% and 60\% accuracy in identifying the quantity and direction of bias present in the AI, respectively) and thus determine that participants were likely both mindful of the potential for bias within any given AI recommendation and to whom that bias was causing unfairness towards. Since our counterbalanced methodology ensured equal participation across treatment orders, we mitigated the confound of simple mindfulness of bias invoking fairer decision--making behaviour within our sample. Therefore, we conclude that the change in behaviour towards more fair decision--making practices must be the result of a conscious and deliberate intent from participants to specifically moderate AI bias.

\section{Discussion}

In our study, we found that the presence of fairness explanations had a slightly positive effect on the performance of decision--making (\textbf{RQ1.a}). In particular, we found that fairness explanations helped participants significantly improve their ability to correctly deny parole to re-offending criminals (PPV), with a negligible impact on overall decision accuracy. This suggests that participants may shift their decision--making behaviour to align more closely with the values encoded by fairness explanations, which could help align distributed decision--making with social values and potentially improve the consistency of decision--making. However, it is currently unknown why this phenomena occurs: our counterbalanced methodological design helps rule out the possibility of our choice of fairness criteria naturally aligning with our participants' prior beliefs and thus confounding the results. In addition, our counterbalanced intervention ordering helps to rule out the possibility that `awareness to the potential for bias' is contributing to its mitigation~\cite{Noon_2018_pointless_diversity_training}. It is possible that, given the attribution of bias to an external entity results in higher rates of perception of the influence of bias in decision--making~\cite{people_see_more_bias_in_algorithms}, participants were more inclined to accept and manoeuvre to counter the influence of bias coming from the AI within a team dynamic. In addition, our use of systematically balanced AI--assistance and fairness explanations meant that no utility was afforded to participants through interaction with the explanations alone. In light of these factors, it remains unclear why participants exhibit significantly fairer behaviour (across multiple metrics) whilst using fairness explanations. Nevertheless, designers of future fairness explanations should be mindful that their designs and choice of fairness criteria are likely to influence the behaviour of those working alongside AI--assistance. Future work should explore how Fairness Explanations affect user behaviour in more depth, especially where fairness criteria may be more difficult to satisfy or where they may contend with current social norms.

Regarding the nature of biases explored in our study: prior work~\cite{dressel_accuracy_2018, chiang} provides no comparable data on pre--existing sex biases in decision-making performance. As a result, we contribute data evidencing baseline sex--biases in the domain of parole decision-making, in addition to the impact of AI--assistance and fairness explained AI--assistance. To this end, participants typically showcased lower deltas in sex biases when compared to race biases, indicating that race--based discrimination may be both more pronounced and yet more difficult to suppress than sex--based discrimination, despite our fairness explanations being designed with the express intent of highlighting race biases in AI recommendations and also with participants never being made aware of the potential for AI to discriminate against sex--groups within this study.

Whilst the majority of our statistical tests for fairness found significant reductions in bias in our sample between the \textbf{Human} and \textbf{Human--FairXAI} conditions, there were three instances where we found no significance after an aggressive adjustment to the $\alpha$ threshold accounting for multiple hypothesis testing. These were: PPV (Sex), PPV (Race) and FNR (Race). A possible reason for this lack of significance could lie in the difficulty or even unwillingness of participants to meet the conditions of the metrics.  In particular, participants in our study were found to have among the strongest initial biases according to these metrics across bias axes and treatment arms and although the reduction of bias means across treatments suggests that participant fairness may have been improving, the high standard deviations suggest that this was perhaps either more difficult to achieve or less well--aligned with how participants were willing to change their behaviour to promote fairness. Future work should seek to explore whether tensions between participant perceptions of fairness and values encoded through fairness criteria play a role on the effectiveness of Fairness Explanations to promote fairer decision--making behaviour.

Whilst we were unable to establish any significant influence of fairness explanations on reducing participant automation bias (\textbf{RQ1.b}), our data suggests that they may be useful in helping to suppress detrimental over- and under-reliance on AI-recommendations~\cite{cabitza_over_under_reliance}. The lack of significance could be a result of variability introduced through our counterbalanced AI--assistance that could make the AI--assistance seem aleatory. When taken in consideration with their bias suppressing effects, fairness explanations may help in mitigating the propensity for immediate and future harm posed by humans learning prejudiced practices from working alongside biased AI~\cite{humans_inherit_bias}.

Finally, whilst our visual fairness explanation proved successful in reducing participant social biases, we found its presence to have an adverse effect on the usability of our decision-making platform (\textbf{RQ1.c}). As a result, the benefits of fairness explanations and their ability to improve utility and reduce bias must be weighed in contention with system usability. Whilst reduced usability was expected given the complexity of the information conveyed through the interactive explanations, the loss in usability did not negate the intended purpose of the explanations in raising awareness of bias. Furthermore, though we observed reduced usability, our fairness explanations still received a marginally highly acceptable SUS score of 65~\cite{SUS_SCORE_MEANING}. Nevertheless, these findings serve as baseline figures for the research community to improve upon in future research exploring the design of usable fairness explanations. The reduction of usability may, in part, be influenced by complexity in the visualisation design: designing a user--centered visual explanation from the ground up that take into account the varying degrees of digital literacy of end--users~\cite{user_centered_design_digital_skills,user_centered_design_recommender_system} may help improve usability for future fairness explanations.

\subsection{Limitations \& Future work}
We faced several factors that limited the generalisability of our work. Since this study only explored one decision--domain with non--expert, US--based participants, it is unclear whether the benefits of fairness explanations persist across multiple domains, or indeed with expert user groups. Future work can address this by adapting fairness explanations for use across multiple domains and assessing them with both non--expert and expert end--users. To reduce complexity within this study, our fairness explanations were designed to convey just one type of unfairness to participants, namely discrimination in denying parole. Since fairness is often described as a subjective phenomena, future work ought to explore how different fairness metrics used in the design of fairness explanations can influences participant behaviour. Moreover, whilst we contribute a potential fairness explanation design within this study, our aim was primarily to explore the decision--making phenomena pertaining to the presence of such an explanation. As a result, there is plenty of room for future work to explore alternative designs. Finally, since both AI assistance and fairness explanations were simulated within this study and did not provide users with additional utility, it is unclear whether the same effects will be observed when participant work alongside highly performant AI assistance. As such, we plan to conduct future research to explore how AI performance impacts the effectiveness of fairness explanations in reducing human--AI bias.

We acknowledge our recruitment conditions serve as potential limitations to the generalisability of our study across cultural domains and that participants from different geo--political contexts may respond differently to visual fairness explanations. We included additional conditions to balance the sample equally based on the race and sex groups of the participants. We set this condition so that the demographics of our participant pool reflected those found in the COMPAS dataset and to remove gender and racial bias in our sampling.

Finally, we acknowledge that our choice to not reduce the scope of real-world decision tasks performed alongside AI judicial risk-assessments from decisions, to predictions (unlike in contemporary studies~\cite{chiang,dressel_accuracy_2018}), likely increased the level of noise observed from participant responses and thus reduced power, yet despite this, our results showcase a clear, strong and significant reduction in bias (and increase in performance) when participants made decisions alongside fairness explanations (Section~\ref{sec:results}).

\section{Conclusion}

This paper presented the design, development and evaluation ($n=48$) of a visual fairness explanation technique for use in conveying bias in AI behaviour to human decision--makers working in tandem with AI. We found that fairness-explained AI assistance resulted in significantly fairer decision-making behaviour than unexplained AI-assistance or independent human decision--making (\textbf{RQ1.a}). Furthermore, we found that the presence of fairness explanations significantly improved combined PPV decision performance over unexplained and independent decision--making (\textbf{RQ1.a}). However, our results showed that the introduction of visual fairness explanations did not significantly reduce automation bias (\textbf{RQ1.b}) and further, that our visual fairness explanations significantly reduced the usability of our online decision--making platform (\textbf{RQ1.c}). This work thus lays the foundation for future research to explore alternative techniques of surfacing fairness--centric explanations for bias within AI models working alongside humans in hybrid team dynamics.

\begin{acks}
The main author is funded by the ESRC Wales Doctoral Training Partnership. Work part funded by the European Union. Views and opinions expressed are however those of the author(s) only and do not necessarily reflect those of the European Union or the European Health and Digital Executive Agency (HaDEA). Neither the European Union nor the granting authority can be held responsible for them. Grant Agreement no. 101120763 - TANGO. For the purpose of Open Access, the author has applied a CC BY licence to any Author Accepted Manuscript (AAM) version arising from this submission. We would like to thank all who participated in our pilot and lab-studies for their help in compiling this work.
\end{acks}

\newpage

\subsection{Decision Tasks}
\begin{figure}[h]
    \centering
    \begin{subfigure}[b]{0.45\linewidth}
        \centering
        \includegraphics[width=\linewidth]{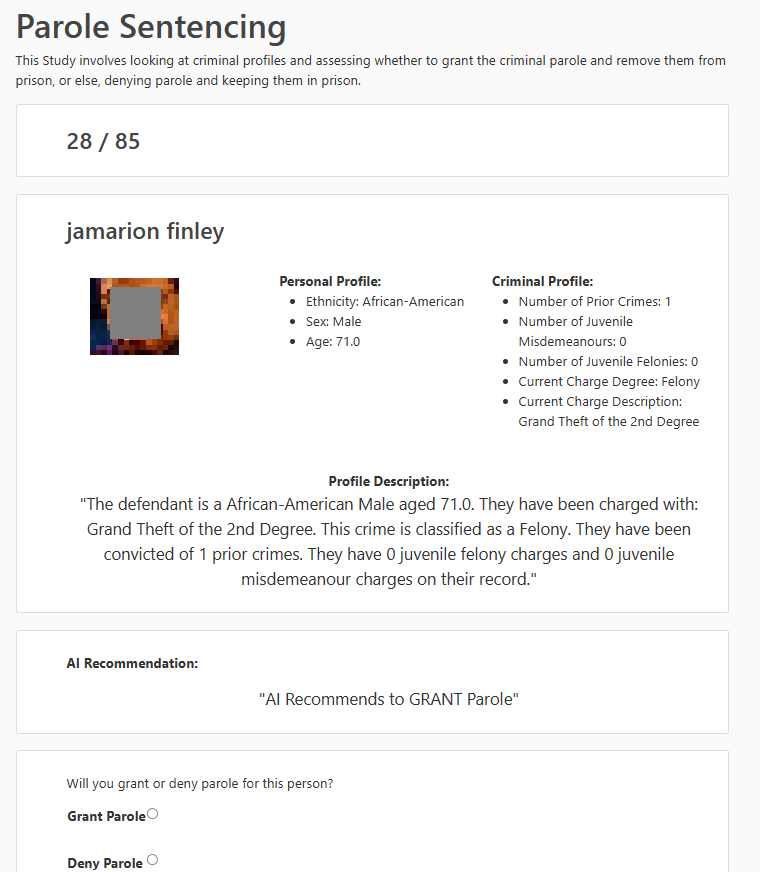}
        \caption{An example of a Human--AI decision task.}
        \label{appx:human_ai_decision_task}
    \end{subfigure}
    \hfill
    \begin{subfigure}[b]{0.45\linewidth}
        \centering
        \includegraphics[width=\linewidth]{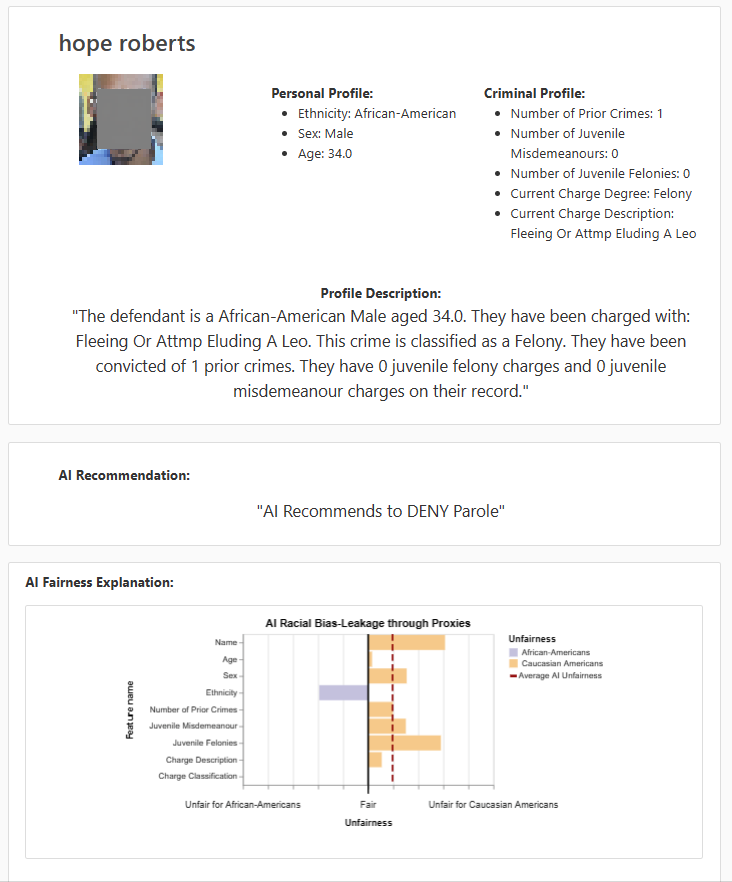}
        \caption{An example of a Human--FairXAI decision task.}
        \label{appx:human_xai_decision_task}
    \end{subfigure}
    \caption{Comparison of Human--AI and Human--FairXAI decision tasks.}
    \label{fig:comparison_decision_tasks}
\end{figure}

For a visual representation of \textbf{Human--AI} and \textbf{Human--FairXAI} based decision tasks, please refer to Figures~\ref{appx:human_ai_decision_task} and~\ref{appx:human_xai_decision_task}

\subsection{Fair Experimental Structure}
\label{appendix: fair_methodology}
\begin{figure}[p]
    \centering
    \adjustbox{rotate=90, width=0.95\textheight}{\includegraphics{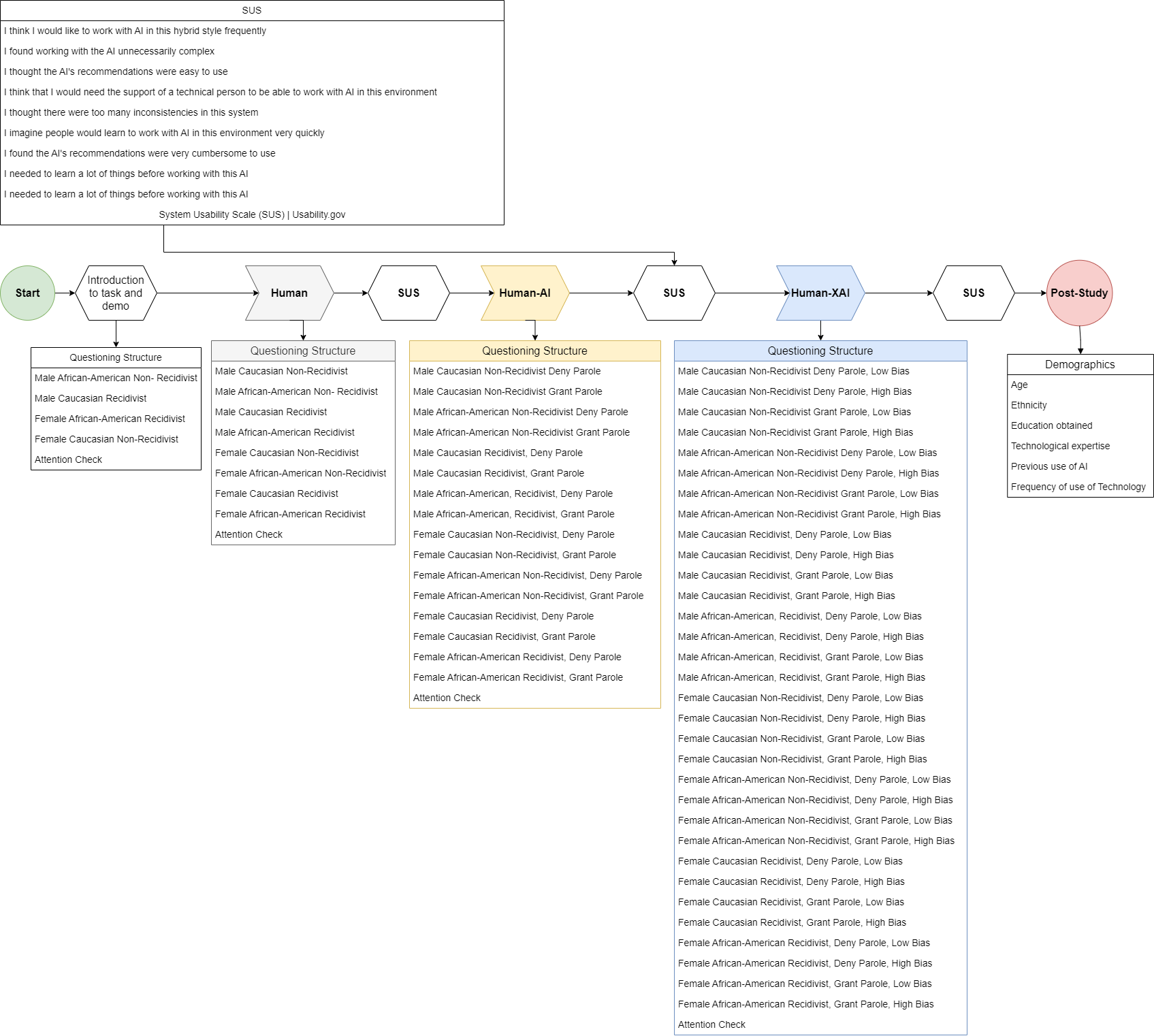}}
    \caption{Complete structure of experimental procedure for Group A, depicting complete coverage of all possible combinations of variables to ensure that no systematic bias was introduced into our sample. Question orderings were randomised in the study, but are laid in sequence here to aid comprehension.}
    \label{appx:study_methodology}
\end{figure}

Please refer to Figure~\ref{appx:study_methodology} for a more in--depth view of our experimental methodology.

\pagebreak
\section{More detail on Results}
\subsection{Accuracy Performances of Participants}

\begin{figure}[h!]
    \centering
    \begin{subfigure}[b]{0.49\linewidth}  
        \centering
        \includegraphics[width=\linewidth]{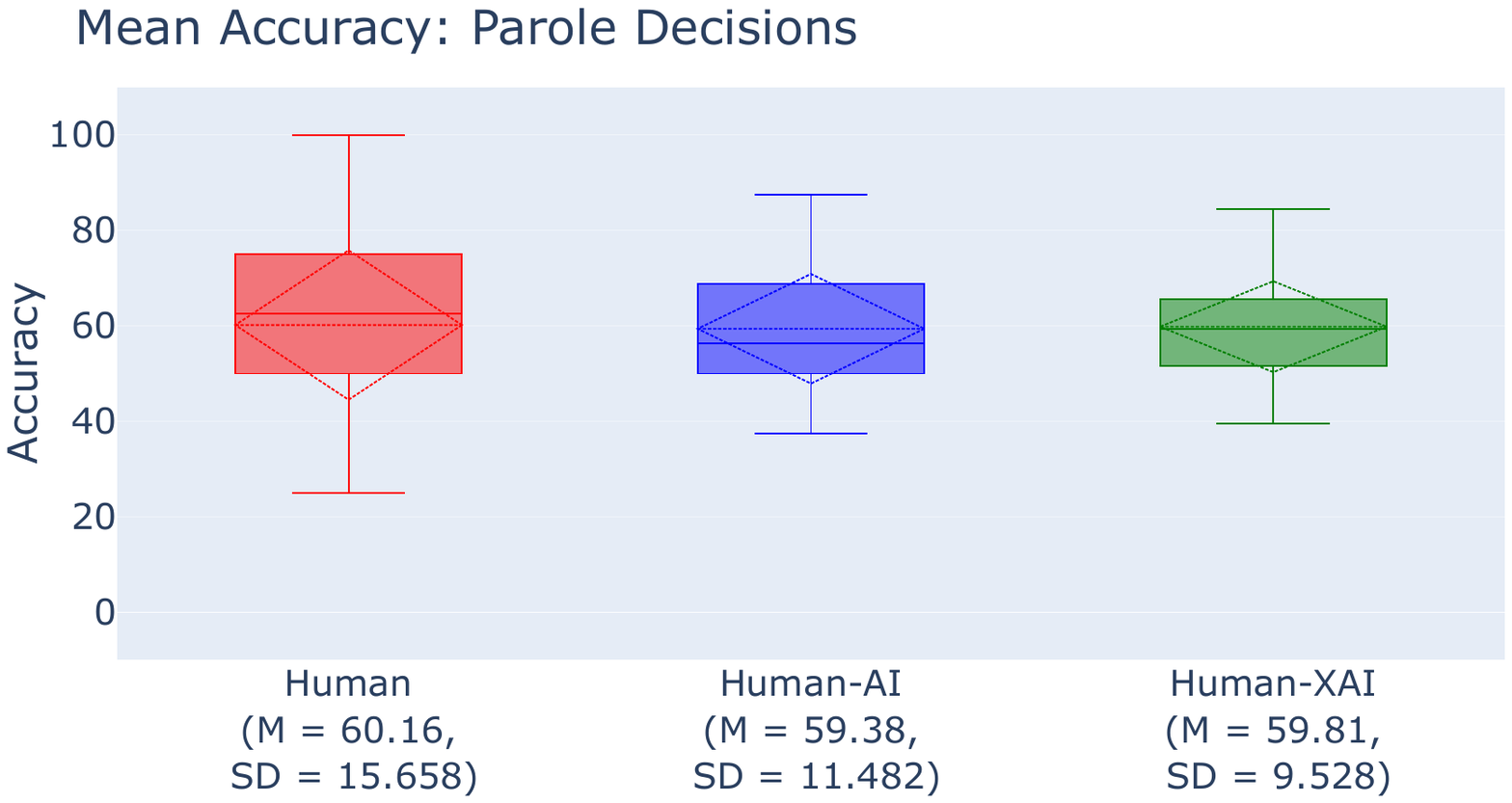}
                    \vspace{-1.5cm}  
        \caption{Participant Mean Accuracy in Parole Decisions}
        \label{fig:mean-accuracy}
        \Description{This chart shows a slight reduction in participant accuracy means across conditions. The full data is reported in Table~\ref{table:fairness} }
    \end{subfigure}
    \hfill
    \begin{subfigure}[b]{0.49\linewidth}  
        \centering
        \includegraphics[width=\linewidth]{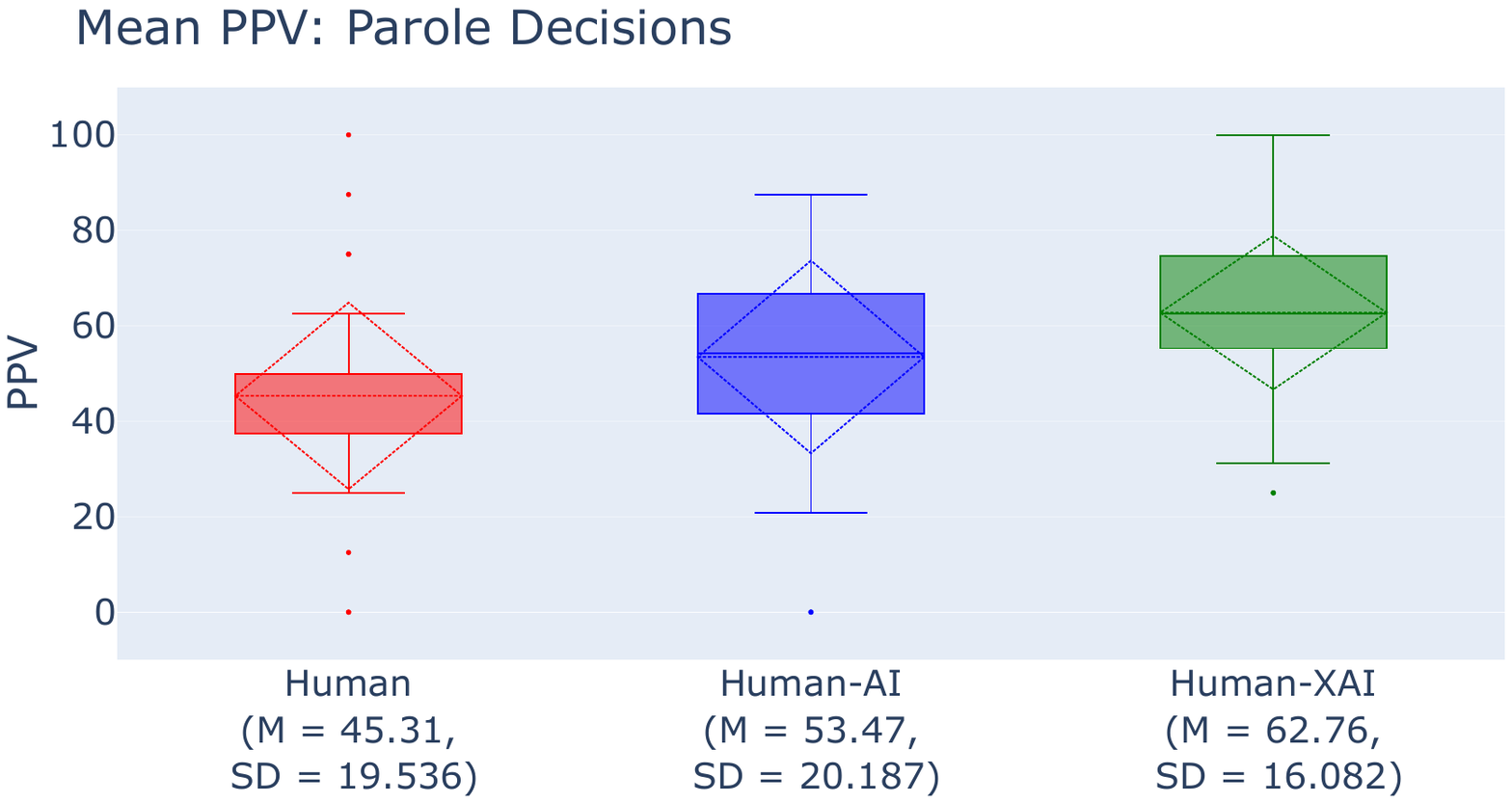}
                            \vspace{-1.5cm}  

        \caption{Participant Mean PPV in Parole Decisions}
        \label{fig:mean-ppv}
                \Description{This chart shows a large increase in participant PPV means across conditions. The full data is reported in Table~\ref{table:fairness} }
    \end{subfigure}
    \caption{Participant accuracy and PPV ratings when making parole decisions across experimental conditions}
    \label{fig:mean_accuracy_ppv_side_by_side}
\end{figure}
\pagebreak

\subsection{Confidence}
\label{appx:confidence}
\begin{figure}[h]
    \centering
    \includegraphics[width=0.45\linewidth]{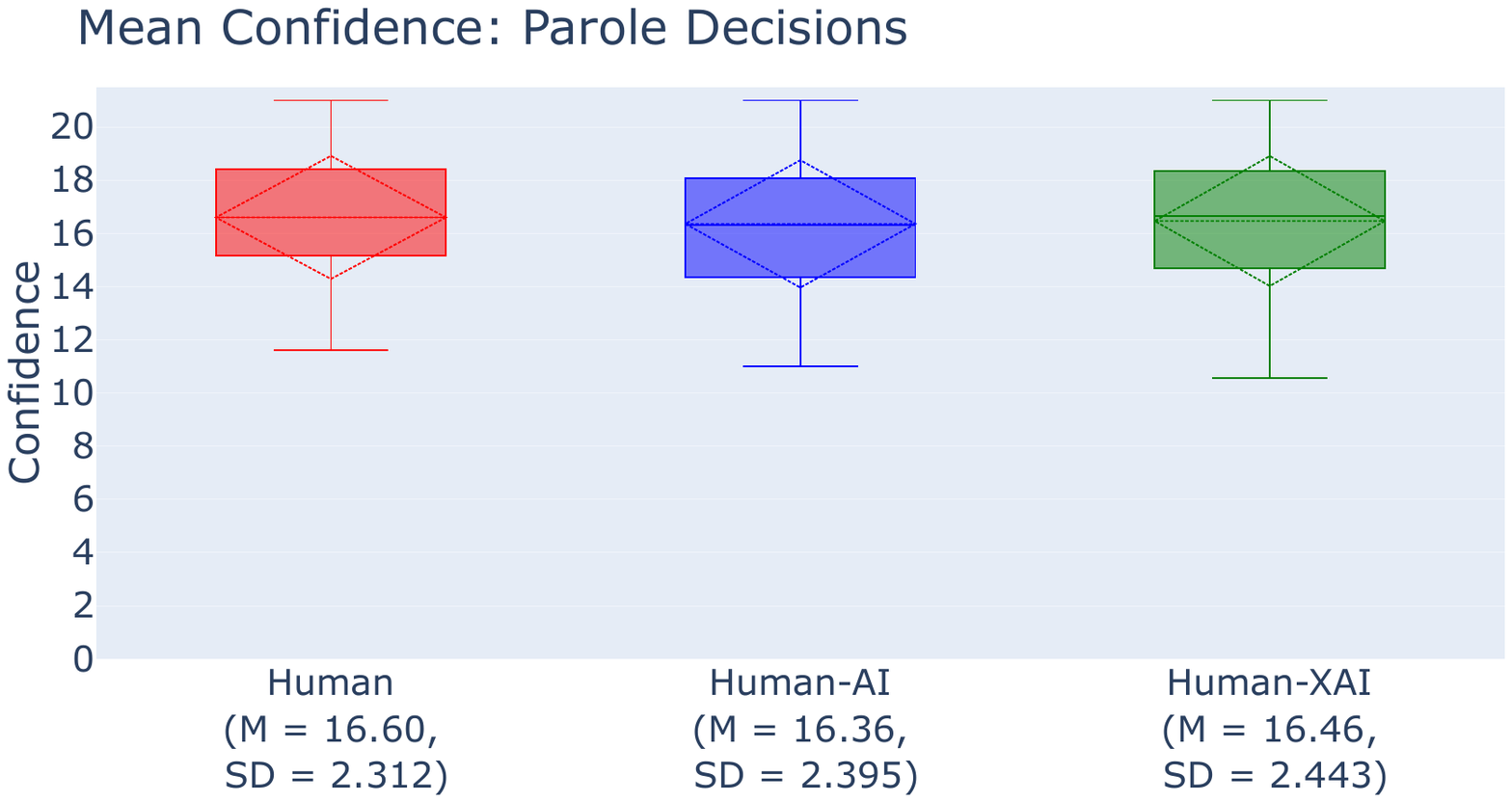}
    \vspace{-1.5cm}
    \caption{Participant Confidence rates in relation to their parole decisions}
    \label{fig:mean_confidence}
    \Description{This figure shows participants confidence in their parole decsions across experimental conditions. All three conditions have very similar mean confidence values.}
\end{figure}
\newpage
\subsection{Automation Bias}
\label{appx:automation_bias}
\begin{figure}[h]
    \centering
    \includegraphics[width=0.45\linewidth]{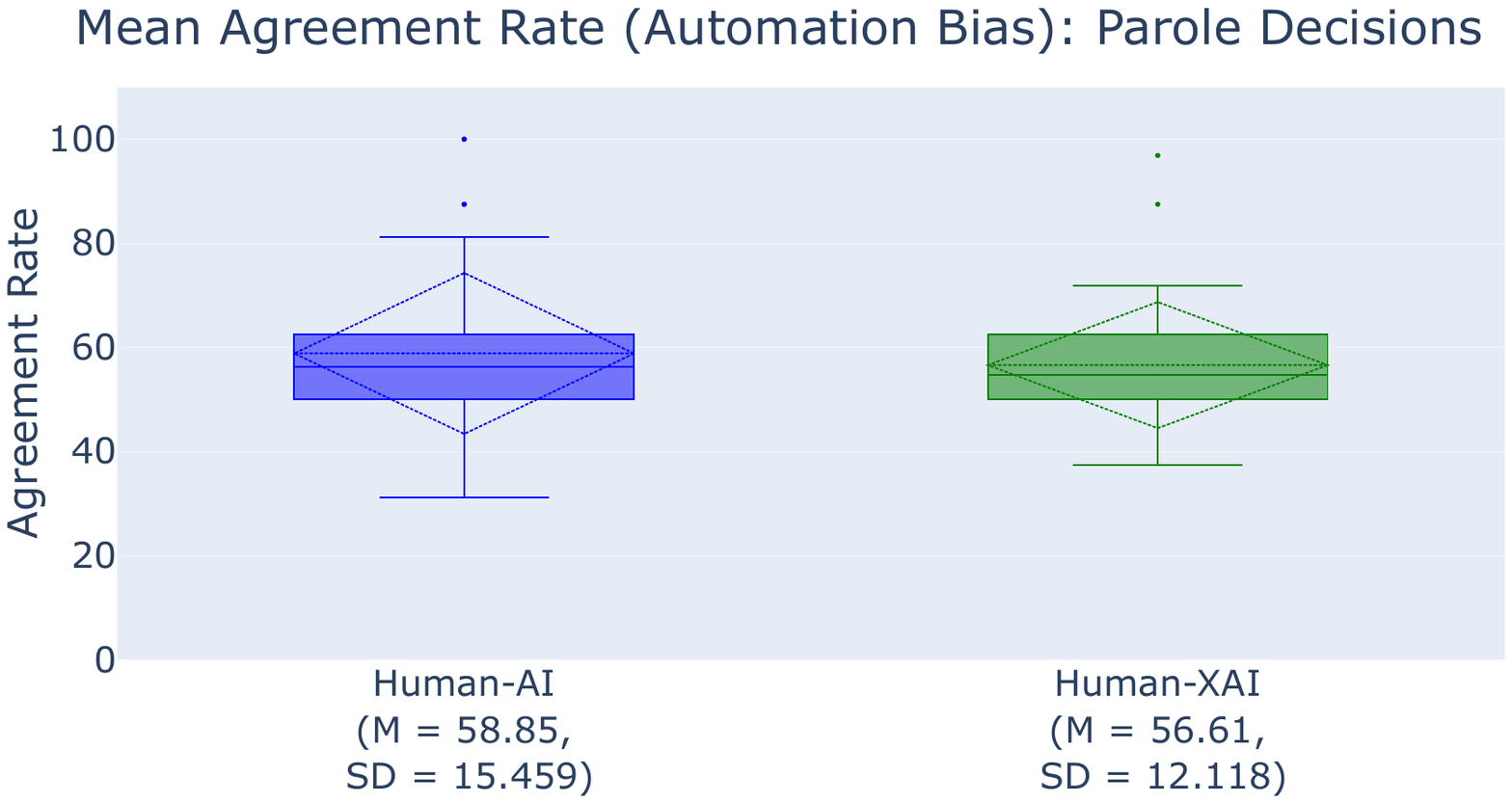}
    \vspace{-1.5cm}
    \caption{Participant automation bias measured by the agreement rate of participants with AI recommendations}
    \label{fig:automation_bias_mean}
    \Description{This chart shows the participants' automation bias via a visualisation of their agreement rates with AI recommendations. The chart shows a slight decline in agreement rates in the Human-FairXAI condition, over the Human-AI condition.}
\end{figure}
\newpage
\subsection{System Usability Scores}
\label{appx:sus_scores}
\begin{figure}[h]
    \centering
    \includegraphics[width=0.45\linewidth]{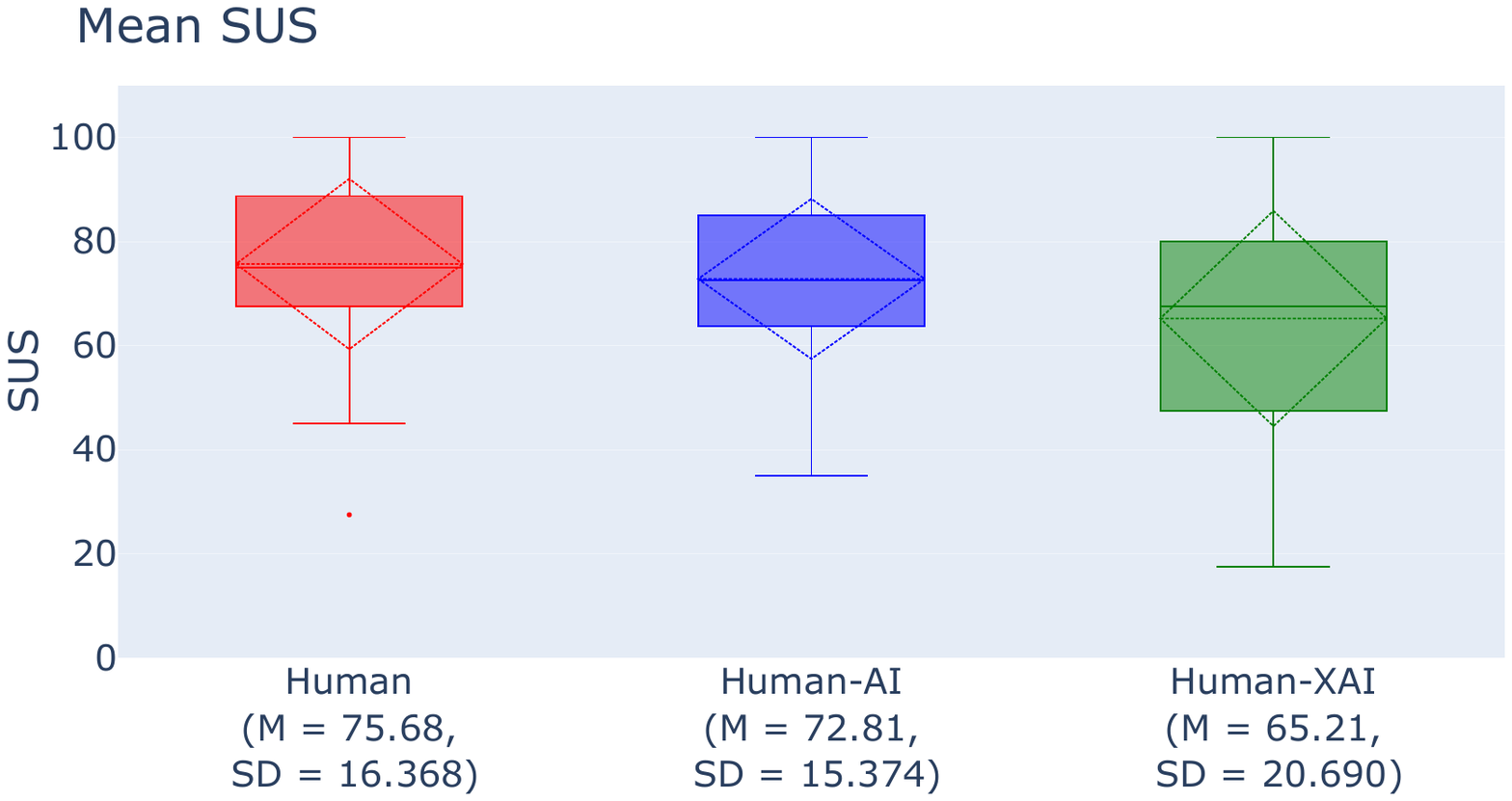}
        \vspace{-1.5cm}

    \caption{Participant SUS scores across conditions}
    \label{fig:sus}
    \Description{This figure shows the declining SUS scores given by participants for the AI and Human-FairXAI conditions.}
\end{figure}

\bibliographystyle{ACM-Reference-Format}
\bibliography{references}

\end{document}